\documentclass[11pt]{article}
\usepackage[centertags]{amsmath}
\usepackage{amssymb}
\usepackage{color} 
\usepackage{rotating}
\usepackage[T1]{fontenc}
\usepackage[utf8]{inputenc}
\usepackage{authblk}
\usepackage{enumitem}
\usepackage{url}

\usepackage{graphicx}
\usepackage{caption}
\usepackage{subcaption}

\begin{document}

\title{A hybrid model of kernel density estimation and quantile regression for GEFCom2014 probabilistic load forecasting}
\author[1]{Stephen Haben\thanks{Stephen.Haben@maths.ox.ac.uk}}
\author[2]{Georgios Giasemidis\thanks{G.Giasemidis@reading.ac.uk}}
\affil[1]{Mathematical Institute, University of Oxford, OX2 6GG, UK}
\affil[2]{CountingLab Ltd, Reading, RG2 8EF, UK}
\date{}
\maketitle

\begin{abstract}
We present a model for generating probabilistic forecasts by combining kernel density estimation (KDE) and quantile regression techniques, as part of the probabilistic load forecasting track of the Global Energy Forecasting Competition 2014. The KDE method is initially implemented with a time-decay parameter. We later improve this method by conditioning on the temperature or the period of the week variables to provide more accurate forecasts. Secondly, we develop a simple but effective quantile regression forecast. The novel aspects of our methodology are two-fold. First, we introduce symmetry into the time-decay parameter of the kernel density estimation based forecast. Secondly we combine three probabilistic forecasts with different weights for different periods of the month. 
\end{abstract}

\section{Introduction}

In this paper we present our methodology used in a winning entry for the probabilistic load forecasting track of the Global Energy Forecasting Competition 2014 (GEFCom2014). The competition consisted of twelve weekly tasks which require using historical data for the estimation of $99$ quantiles ($0.01, 0.02, ..., 0.99$) for each hour of the following month. Each forecast is evaluated using the pinball function. 
For further details on the competition structure and the data the interested reader should refer to the GEFCom2014 introduction paper \cite{GefComIntro2015}.
In Section \ref{PrelimAnalysis} we present a preliminary analysis of the data that motivates the development of the main forecasting methods introduced in Section \ref{Methods}. In Section \ref{taskOutlines} we give a short description of our submissions in chronological order to explain the reasoning behind the chosen forecasts and the developments of the subsequent forecasts. We present an overall view of the results and conclude in Section \ref{Discussion} with a discussion, lessons learned and future work.

\section{Preliminary Analysis} \label{PrelimAnalysis}

We start by performing a preliminary analysis to determine our initial forecast methods. We first tested the competition's initial historical data set to confirm that load and temperature are strongly correlated, as shown in other studies \cite{Charlton2012}, see also the GEFCom2014 introduction paper \cite{GefComIntro2015} for the time-series plots of the data. This motivates the development of our kernel density estimation method conditional on the temperature (see Section \ref{sec:CKDT}). We also found that all the weather stations were strongly correlated with each other and the load data. Hence as an initial estimate of the temperature we simply took an average over all $25$ stations.

The load data has strong daily, weekly and yearly seasonalities as well as trends \cite{GefComIntro2015}. A visual analysis of the load data showed that certain hours of the day exhibited strong bi-annual seasonalities (such as $11$pm) whereas others did not (e.g. $3$pm). This could be due to heating and cooling appliances being employed through the seasons. This inspires our choice of biannual model in the quantile regression based forecast (see Section \ref{QR}). Consideration of the autocorrelation and partial autocorrelation plots confirmed the presence of the weekly and daily periodicities. Our forecasts described in the following section are influenced with this periodicity in mind. 

\section{Methododology} \label{Methods}

In this section we present the main methods implemented for the competitive tasks of the competition.

\subsection{Kernel Density Estimation (KDE)} \label{KDE}

Many of the methods we employ are non-parametric kernel density based estimates and similar to those as presented in \cite{Jeon2012} for probabilistic wind forecasting and \cite{Arora2014} for household-level probabilistic load forecasting. This method is motivated by the strong weekly correlations in the data. A simple kernel density estimate produces an estimate of the probability distribution function $f(X)$ of the load $X$ (at a particular future time period) using past hourly observations $\{X_i\}$ (assuming $i=1$ is beginning of historical load data: $1^{st}$ Jan 2005.). It is given by
\begin{equation}
f(X)=\frac{1}{nh_x}\sum_{i=1}^n K \left( \frac{X-X_i}{h_x} \right),
\label{eq:KDE}
\end{equation}
where $h_x$ is the load bandwidth. 
We use a Gaussian kernel function, $K(\bullet)$, for all our kernel based forecast methods. 
Our first method is a KDE with a time decay parameter, $0<\lambda\leq 1$. The role of the decay parameter is to give higher weight to more recent observations. To forecast day $D$ of the week, $D = 1, 2, \dots, 7,$ at hour $h$, $h = 1, 2, \ldots, 24$, we applied a KDE on all historical observations of the same day $D$ and hour $h$. This method only considers observations belonging to the same hourly period of the week, denoted by $w$, $w = 1, \ldots, 168$, and we refer to it as \textit{KDE-W}. This can be expressed as
\begin{equation}
f(X)=\frac{1}{nh_x}\sum_{\substack{ i=1 \\ \{ i\!\!\!\!\!\mod \!s = w \} }}^n \frac{\lambda ^{\alpha(i)}}{\sum_{\substack{i=1 \\ \{i\!\!\!\!\!\mod \!s = w\} }}^n\lambda^{\alpha(i)}} K \left( \frac{X-X_i}{h_x} \right).
\label{KDEW}
\end{equation}
The parameter $s = 168$ is the number of forecasting hours in a week and $\alpha(i)$ is a periodic function given by
\footnote{The careful reader should note that the formula \eqref{expoKDEW} might need a further correction by one when $D$ is in a leap year. However this does not affect our results, since we did not forecast leap years. Additionally such an error would have a negligible effect in the weight.}
%
\begin{eqnarray}
\alpha(i) &=& \min \left (|\mathcal{D} - (\mathcal{D}(i)-\mathbf{1}_{A}(i))|, \mathcal{T}(i) - |\mathcal{D} - \mathcal{D}(i)|\right ),
\label{expoKDEW}
\end{eqnarray}
%
where $\mathcal{D}(i) = 1, 2, \ldots, \mathcal{T}(i)$ is the day of the year (consisting of $\mathcal{T}(i)$  days) corresponding to the historical data $X_i$ and $\mathcal{D}$ is the day of the year corresponding to the forecasted day. To correct for leap years we use an indicator function $\mathbf{1}_{A}(i)$ where $A=\{i |  \mathcal{D}(i)>28 \text{ and } \mathcal{T}(i)=366 \}$. Expression \eqref{expoKDEW} is simply a periodic absolute value function with annual period, whose minimum values occur annually on the same dates as the forecasted day.

This method is similar to the one presented in \cite{Arora2014}. However the new feature is the half-yearly symmetry of the time-decay exponential \eqref{expoKDEW}. Since there is an annual periodicity in the load we incorporated it into the time-decay parameter such that observations during similar days of the year influence the forecast more than other, less relevant observations. The decay parameter also helps us to take into account the non-stationary behaviour of demand. This method performed better compared to a similar KDE-W using only a simple monotonically decreasing time-decay parameter across the year. 
The model parameters were generated using cross-validation on the month prior to the forecasting month. To find the optimal bandwidth, $h_x$, we used the \textit{fminbnd} function from the optimisation toolbox in Matlab. For the time-decay parameter $\lambda$ we considered different values between $0.92$ and $1$ with $0.01$ increments\footnote{The time-decay parameter must be in the interval $(0,1]$, the smaller the value the fewer historical observations which have significant influence on the final forecast. After testing over several tasks we found that the decay parameter is bounded below by $0.92$.}.

The kernel density based estimate has been used as a benchmark in probabilistic forecast methods applied to household level electricity demand. It serves as a useful starting point for our forecasts \cite{Arora2014}. The method has the advantage of being quicker to implement than more complicated kernel based methods, such as the conditional kernel density estimate on independent parameters, which we introduce in the next sections. 

\subsection{Conditional Kernel Density Estimate on Period of Week (CKD-W)} \label{sec:CKDW}


A KDE forecast conditional on the period of the week, denoted by $w$, $w = 1, \ldots, 168$, (CKD-W) \cite{Arora2014} gives a higher weight to observations from similar hourly periods of the week and can be represented as
\begin{equation}
f(X|w)=\sum_{i=1}^{n}\frac{\lambda^{\alpha(i)} K((w_i-w)/h_w)}{\sum_{i=1}^{n}\lambda^{\alpha(i)} K((w_i-w)/h_w)}K \left( \frac{X-X_i}{h_x} \right)
\label{CKDW}
\end{equation}
where $\alpha(i)$ is defined in \eqref{expoKDEW}.

This method is similar to the one presented in \cite{Arora2014}. However the new feature is the half-yearly symmetric time-decay exponential \eqref{expoKDEW} which is justified by the yearly periodicity of the load as explained in the previous section. 

The validation process can be computationally very expensive, especially while searching for multiple optimised parameters (here there are three parameters, the bandwidths for load and week period variables, and the time decay). 
In particular, despite using the Matlab parallelisation toobox, executing this method on our (conventional) machines\footnote{All forecasts were executed on a machine with Intel Core i7-361QM Quad-Core Processor @ 2.30GHz and 16GB of memory.},
required more than a day to complete, which is not practical given the weekly constraints of the competition. In an attempt to reduce the computational cost, we reduced the number of historical observation and the length of the validation period. We only used observations starting from January of $2008$ and we cross-validated our parameters using only 
one week from the validation month\footnote{Initially we used the first week, but later we used the last week from the validation month because it is closer to the period to be forecasted. However the improvement was minor.}.

For the optimisation of the bandwidths we used the \textit{fminsearch} function (implementing a $\log$ transformation to ensure that we only model for positive values) from the optimisation toolbox in Matlab. For the time-decay parameter we looped over different values of $\lambda$ between 0.92 and 1 with 0.01 increments. At the final stages of the competition we used the \textit{fminsearchbnd} function\footnote{\url{http://www.mathworks.com/matlabcentral/fileexchange/8277-fminsearchbnd--fminsearchcon}.}
for parameter optimisation, which improves both the computational time and the accuracy. 
We call this implementation of the method CKD-W2, see also Section \ref{taskOutlines}.

\subsection{Conditional Kernel Density Estimate on Temperature Forecast (CKD-T)}
\label{sec:CKDT}
Weather information is particularly useful for an accurate load forecast (among many references in the literature see \cite{Jeon2012} in the context of CKD methods, and also a winning entry of GEFCom2012 \cite{Charlton2012}). For this reason we implemented a KDE method conditional on the temperature (CKD-T). We take the explanatory variable to be the mean hourly temperature $T$ from the 25 weather substations. The conditional probability density is given by
\begin{equation}
f(X|T)=\sum_{i \in \mathcal{A}}\frac{ K((T_i-T)/h_T)}{\sum_{i \in \mathcal{A}}K((T_i-T)/h_T)}K \left( \frac{X-X_i}{h_x} \right),
\label{CKDT}
\end{equation}
where $h_T$ is the bandwidth of the temperature kernel and $T_i$ is the temperature corresponding to the same hour $h$ and day $d$ as the load $X_i$. The index subset $\mathcal{A}$ consists of indices at time $h$ and days $d-5,\ldots,d,\ldots,d+5$ of all previous years. The formula \eqref{CKDT} does not include a time-decay parameter since we assume the temperature is the main driver of seasonality. Thus we do not include a decay parameter which would increase the computational expense for very little gain. For parameter optimisation we used the \textit{fminsearch} function, implementing a $\log$ transformation as with the CKD-W forecast.

Since temperature forecasts are inaccurate beyond a few days this method was only implemented for the first day of a task. 
As we will shortly describe in Section \ref{Sec:Comb}, the remaining days of a task are forecasted using a weighted combination of CKD-W and a quantile forecast, introduced in Section \ref{QR}.

\subsubsection{Temperature Forecast} \label{TempForecastStuff}

The CKD-T method requires a forecast of the mean temperature in order to create a load forecast. We follow a simple autoregression forecast method, similar to that presented in \cite{Liu2014}. The model was chosen for its simplicity. In addition, temperature can change rapidly within a couple of days, and without more data (such as wind speeds and direction), and the access to complicated numerical weather prediction software we decided a simple model is appropriate for our uses. The model consists of a trend, seasonalities (both diurnal and yearly) and lagged temperature variables. We model the temperature $T_j$ at timestep $j$ as
\begin{equation}
T_j=\beta_0+\beta_1 j+S_j^d+S_j^a+\sum_{k=1}^{25} \alpha_k T_{j-k}.
\label{eq:TempForecast}
\end{equation}
The diurnal seasonal terms are described by
\begin{equation}
S_j^d=\sum_{p=1}^P\left (\gamma_p \sin\left( 2\pi p \frac{d(j)}{24} \right)+ \delta_p \cos \left( 2\pi p \frac{d(j)}{24} \right) \right ),
\label{eq:}
\end{equation}
where $\gamma_p, \delta_p$ are Fourier coefficients  (with $P=4$) and $d(j)= j \mod 24$ is the conversion to the hour of the day.
The yearly seasonal terms are modelled by
\begin{equation}
S_j^a=\sum_{m=1}^M\psi_m \sin\left( 2\pi m \frac{(f(j) + \phi)}{365} \right),
\label{eq:}
\end{equation}
where $\psi_m$,  $m=1, 2, ..., M$ and $M=3$, are the coefficients and $f(j)= j/24$. The method slightly differs from that in \cite{Liu2014} which uses $f(j)=\lfloor j/24 \rfloor$ (the day of the data).  The shift $\phi$ ensures the period terms match the period of the data as optimally as possible. The value $\phi=-85$ was chosen such that the mean absolute percentage error (MAPE) is minimised. We set $j=0$ for the start of data at midnight on $1^{st}$ January $2005$. The final terms of equation (\ref{eq:TempForecast}) are the lags. By consideration of the autocorrelation, we checked the potential number of lag terms to use and found that the previous $25$ hours gave the minimum MAPE for day ahead and month ahead temperature forecasts over November 2009 (a preliminary task). The values of $M, P$ and $\phi$ were all chosen by cross validation over the month of November $2009$. The coefficients $\beta_0, \beta_1, \gamma_p, \delta_p$ and $\psi$ were found via the linear regression function in Matlab, \textit{regress}.

We attempted to select the most representative and accurate weather stations to improve the day ahead CKD-T forecast. We chose groups of three and six weather stations which gave the best MAPE for a day ahead temperature forecast. Using the average temperature from these stations in (\ref{eq:TempForecast}) did not provide a consistent improvement in the pinball scores. Hence we only used the mean over all weather stations for the CKD-T day ahead forecasts.

\subsection{Quantile Regression (QR)} 
\label{QR}

The quantile regression is a generalisation of standard regression where each quantile is found through the minimisation of a linear model to historical observations according to a loss function \cite{Koenker1978}. Suppose we have a model of the demand, at time $t=1, ..., n$ given by $f(\mathbf{U}_t, \boldsymbol\beta)$, where $\mathbf{U}_t$ are the independent variables and $\boldsymbol\beta$ are the unknown model parameters. Also suppose we have observations of the load $y_t$ at the same times $t=1, ..., n$. Then for a given quantile $q$ the aim is to find the parameters $\boldsymbol\beta_q$ given by
\begin{equation}
\boldsymbol\beta_{q}=\underset{\boldsymbol\beta}{\operatorname{argmin}}  \sum_{t=1}^{n}\rho_{q}(y_{t}-f(\mathbf{U}_t, \boldsymbol\beta)), 
\label{eq:Quantileregression}
\end{equation}
where $\rho(\bullet)$ is the loss function given by
\begin{equation}
\rho_{q}(z)=|z(q-\mathbf{1}_{(z<0)})|,
\label{eq:lossfunction}
\end{equation}
where $\mathbf{1}_{(z<0)}$ is the indicator function.  We created a simple linear function, for each hour of the day separately, based on only trend and seasonal terms. For each daily hour on day $k$  (with $k=1$ meaning $1^{st}$ Jan $2005$) of the data set, we define our model by
\begin{equation}
L_k=a_0+a_1 k+\sum_{p=1}^2 b_p \sin\left( 2\pi p \frac{(k +\phi_1)}{365} \right)+ \sum_{m=1}^2 c_m \sin \left( 2\pi m \frac{(k+\phi_2)}{365} \right).
\label{eq:}
\end{equation}
The first shift term is chosen $\phi_1=-111$, by minimising the MAPE, and the second shift is $\phi_2=\phi_1-364/2$. The double seasonality offset term was used because of the double yearly period discovered in the load for some hours of the day. The coefficients $a_0, a_1, b_1, b_2, c_1, c_2$ are found for each quantile forecasted via a simple linear programming method. We implemented this using \textit{optimset} function in Matlab utilising the Simplex algorithm option. To reduce computational cost we only used $500$ days of history to find the parameters. 
Once the quantile forecasts were found we resorted them to ensure there was no crossing of the quantiles  \cite{Chernozhukov2010}. 

\subsection{Mixed Forecasts and Hybrid Forecasts} \label{Sec:Comb}

Each of the main forecasts presented had different performance for different forecast horizons. For this reason we created new forecasts which were mixes of our main methods based on their performances over different horizons. We consider two main methods
\begin{itemize}
	\item Mix 1: This is simply the CKD-W forecast but using the CKD-T forecast for the first day.
	\item Mix 2: As mix 1 but using the QR forecast from the start of the $8^{th}$ day until the end of the month.
\end{itemize} 

With the success of the mixed forecasts (see Section \ref{taskOutlines}) we also explored combinations of the forecasts. This has been shown to improve the overall forecast accuracy compared to individual forecast methods \cite{Rangan2008}. 
We split the forecast into five different time periods. Period one was simply the first day, period two the rest of the first week, period three the second week, period four the third week and period five the rest of the month. For the first period we simply used the CKD-T which had the best day ahead accuracy of all the forecasts. For each of the other periods we took a weighted average of the quantiles time series in the quantile regression forecast, $F_{\text{QR}}$, and the CKD-W forecast, $F_{\text{CKD-W}}$, 
\begin{equation}
F_{\text{Hybrid}}(\tau) = w(\tau) F_{\text{CKD-W}}(\tau) + (1 - w(\tau)) F_{\text{QR}}(\tau), 
\end{equation}
where $\tau = 2,3,4,5$ is the time period and $0 \leq w(\tau) \leq 1$ is the average optimal weight at time period $\tau$. The optimal weight of each past task is found by searching different weighted combinations of the CKD-W with the quantile regression forecasts for each time-period $\tau > 1$ that minimise the pinball scores. We repeat this process for a number of  past tasks and then take the average optimal weight for each time period.  
We call this forecast the \textit{hybrid forecast}.

\section{Task Submissions and Results} \label{taskOutlines}

\begin{figure*}
	\begin{center}
		\includegraphics[scale = 0.3]{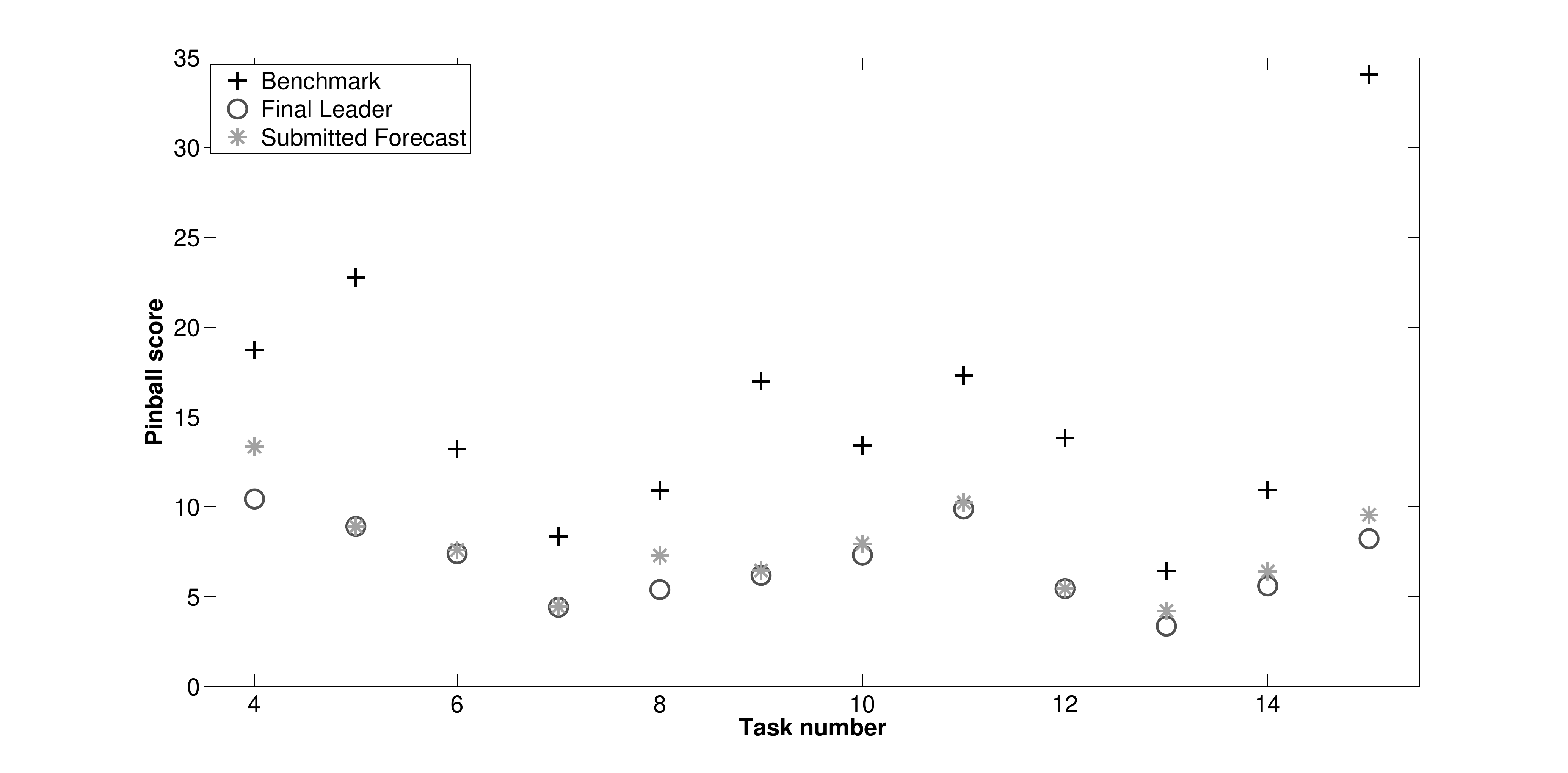} %
		\caption{Pinball scores of our submitted forecast, the benchmark and the final leader.}
		\label{AllScores}
	\end{center}
\end{figure*}

We ranked our forecasts using the scores from prior tasks. We used this to understand which methods to persist with and which ones to reject or adapt. In this section we describe our selection procedure for each task in chronological order to justify our methodology and approach. Figure \ref{AllScores} shows graphically the scores for our best submissions, the benchmark and the top scoring forecast for each task\footnote{Tasks 1 to 3 were trial tasks. We focused on searching for patterns, trends and correlations in the load and temperature data and developing our more sophisticated methods. We submitted simple parametric models and the KDE-W method.}. The plot shows our forecasts performing consistently well in all tasks other than task four and eight as we will explain below. We note that the leader is not the same entrant for each task. The benchmark is simply the previous year's load used for all quantiles. 

In \textbf{tasks 4 and 5}  we implemented the KDE-W method (see Section \ref{KDE}). December $2010$ (task 3) appeared to have unusually low temperatures and since this month was also used for parameters training it could explain the high scores of most entrants of task 4. We note that the simple quantile regression forecast (introduced in task 9) performs very well for this method, scoring 10.36, in fact beating the top entry score. This could be due to being less influenced by the previous, exceptional, month. 

For \textbf{tasks 6 and 7}, we developed the CKD-W method to take into account weekly effects. This was found to perform better than the KDE-W method. We also submitted a CKD-W for \textbf{task 8} but trained the parameters on the same month of the previous year, rather than the previous month of the same year. The data from the previous year would be less recent but likely related to the current month's behaviour due to annual periodicity of the load. In addition, data from the previous month had little influence on forecasts of beyond a week so it made sense to attempt to optimize parameters on data available for the entire period. 
Although this method performed better than CKD-W for task 7 the method did not perform as well as expected for the task 8 submission and was abandoned from thereon. 

We found that the CKD-T method, although poor for forecasting the entire month was the most successful method for forecasting a day ahead (see Section \ref{sec:CKDT}). In addition we developed the QR forecast which was performing well, especially at horizons of over a week ahead. Hence, for \textbf{tasks 9, 10 and 11} we implemented our mixed forecasts. Modifying the first day forecast with the CKD-T forecast, to create mix forecast 1, gave us improved forecast for task 9. Further improvements came with mix forecast 2 which was used as our submission for task 10 and 11 (giving us second place in both leaderboards). Further testing of the forecasts on older tasks indeed confirmed the improvement of the methods. Up to this point the mix 2 forecast gave the more consistent best scores for tasks 2 through to 8 with the smallest average pinball score of $ 8.61 $ compared to the next best of the quantile forecast with the CKD-T for the first day of $8.63$ (the benchmark average was $15.28$). This seems to indicate that a major contribution to the improvement came from the quantile regression forecast. 

For \textbf{tasks 12 to 15} we implemented the hybrid forecast. For these tasks we trained the weights using tasks 6 to 11, 2 to 12, 2 to 13 and 2 to 14 respectively.  This forecast performed better for each task compared to our other methods, see also Table \ref*{TableScoresFinal}. For task $15$, we initially attempted to model separately the special days, Christmas Eve, Christmas Day and New Year's day. However we saw no improvement in our forecasts and since these days all occurred on weekends for task 15, we abandoned this idea. The hybrid models were consistently the best model for tasks 12 to 14 with an average pinball score of $5.36$ compared to the next best score of $5.41$ for the CKD-W2 method. However for task $15$ the method did not perform too well, with a pinball score of $9.55$ compared to only $7.844$ for the KDE-W and $8.099$ for the CKD-T (the winning score was $8.229$). In fact the CKD-T method performed surprisingly well for tasks 12 through 14 with an average score of $5.42$ meaning a better score on average than the hybrid forecast for tasks 12 through 15 ($6.089$). This is particularly surprising given the CKD-T method had the worst performance of all methods prior to task 12. This could possibly be the result of relatively stable temperatures for these months.


The final scores were calculated as a weighted average of the percentage improvement relative to the benchmark for each task. Each percentage score was given a weight which increased linearly from the fourth to last task. 
The scores for selected methods (plus, for comparison the scores of the leading submission for each task \footnote{The leading submission is the best submission from all teams for each task. Not to be confused with the submission of the winning team.} and the competition's winning team) are shown in Table \ref{TableScoresFinal}. The larger the score the better the forecast. The hybrid forecast uses the weights used in the final task and therefore is not a completely accurate representation of the actual hybrid forecast score since the data was optimized to the same tasks. However it does show the potential of the method. If we had more time then potentially we could train the weights on a larger sample for each time period by a rolling window rather than, sometimes, less than six tasks. The table shows the improvements made with subsequent tasks. We note, despite the simplicity of the method, the QR forecast is one of the best non-hybrid forecast on average. However on a few tasks this forecast did not perform as well as the CKD-W and CKD-W2 forecasts (tasks 5, 6, 11, 14) and thus a mixed forecast is perhaps a more reliable choice since these methods perform well when QR does and reduce the errors when QR does not perform as well. The better score of CKD-W2 over CKD-W shows the importance of using the best optimization programs for the forecast. 

\begin{table}
\begin{center}
\begin{tabular}{|c|c|c|c||c|c|c|c|c|c|c|}
\hline
Forecast 		 & LS & WT & Actual & Hybrid & QR  &CKD-W2 & Mix 2 & Mix 1 & CKD-W & KDE-W \\
\hline
Score    & 54.2 & 50.8 & 48.5   & 51.4   & 48.7 & 48.7 &48.4 & 47.5 & 47.4 & 44.6  \\
\hline
\end{tabular}
\caption{Weighted average scores of the leading team of each task (``LS''), the competition's winning team (``WT'') and our valid submitted forecasts (``Actual'')  and our main methods described in Section \ref{Methods}.}
\label{TableScoresFinal}
\end{center}
\end{table}

\section{Summary and Discussion} \label{Discussion}

We have described a number of methods for creating probabilistic forecasts and outlined our methodology for adapting these forecasts for each task. We chose and developed these methods based on a number of characteristics including success of the methods in similar applications, their computational simplicity and their versatility in incorporating the periodic nature of the data. We have created several forecasts which perform well and obtain the lead score in a number of tasks. Our forecasts performed consistently well, too. All forecasts beat the benchmark with only three of the twelve submissions not improving on the benchmark by at least $40\%$. Overall we obtained five top two finishes in the twelve tasks, with top position twice and second position on three occasions. This was the second highest top two finishes amongst all final candidates.

There are periodicities in the scores likely due to more variability in load due to heating and cooling. The benchmark and forecast scores are correlated due to this. Very large benchmarks scores are likely due to large differences in weather conditions. In certain tasks (such as 3 and 4) all teams scored poorly. For example, task 3 we found that there were very low temperatures which correlated with large forecast errors on the $14^{th}$ December. The strong correlation between the weather and load demand imply that the biggest single improvement in forecast accuracy will come from better long term weather forecasts. 

Table \ref{TableScoresFinal} illustrates that the hybrid forecast is the best scoring overall. However it is clear that the simple quantile forecast is responsible for much of this improvement with all forecasts using this method scoring very similarly. Although CKD-W2 and QR perform similarly on average, the CKD-W2 only performs better than the quantile regression on a few tasks.  On those tasks the difference is significant and therefore the hybrid forecast reduces this discrepancy.

Further improvements could have been done to further improve the scores. There are a number of changes which may improve the basic forecasts (CKD-W, CKD-T, QR) such as including weekday identifiers or improving the choice of weather stations. However the simplest modification we could make is to improve the weights used in the hybrid forecasts. In particular we could train the weights on a rolling basis from one day to the next. This mean that the most recent (and accurate) weights could be applied, and potentially we could even forecast such weights. In this paper we have reported a simple combination of our two forecast methods to create a hybrid forecast. It has been shown that a simple linear combination is not optimal since, even if the forecasts are properly calibrated, the final forecast will not be \cite{Rangan2008}. Hence we could also consider other methods, for example the beta linear pool method as described in \cite{Rangan2008}. A surprising for us result of the competition was the success of very simple methods. The quantile regression, which only modelled the trend and yearly seasonality, was one of our, and the competitions', best performing forecasts. Such methods could thus be used as benchmarks for more complicated methods.


\bibliography{loadForecastBib}{}
\bibliographystyle{plainurl}


\end{document}